\documentclass[amsmath,amssymb,amsfonts,aps,pra,citeautoscript,10pt,twocolumn]{revtex4-1}
\usepackage{graphicx}
\usepackage{multirow}
\usepackage{booktabs}
\usepackage{array}
\usepackage{amsmath}
\usepackage{url}
\usepackage{threeparttable}
\usepackage{color}
\usepackage[table]{xcolor}

\newcommand{\superscript}[1]{\ensuremath{^{\textrm{#1}}}}

\begin{document}
\title{A comparison of integrated testlet and constructed-response question formats}

\author{Aaron D. Slepkov}
\email{aaronslepkov@trentu.ca}
\affiliation{Trent University, Department of Physics \& Astronomy, Peterborough, ON K9J 7B8, Canada}

\author{Ralph C. Shiell}
\email{ralphshiell@trentu.ca}
\affiliation{Trent University, Department of Physics \& Astronomy, Peterborough, ON K9J 7B8, Canada}

\begin{abstract}
Constructed-response (CR) questions are a mainstay of introductory physics textbooks and exams. However, because of time, cost, and scoring reliability constraints associated with this format, CR questions are being increasingly replaced by multiple-choice (MC) questions in formal exams. The integrated testlet (IT) is a recently-developed question structure designed to provide a proxy of the pedagogical advantages of CR questions while procedurally functioning as set of MC questions. ITs utilize an answer-until-correct response format that provides immediate confirmatory or corrective feedback, and they thus allow not only for the granting of partial credit in cases of initially incorrect reasoning, but furthermore the ability to build cumulative question structures. Here, we report on a study that directly compares the functionality of ITs and CR questions in introductory physics exams. To do this, CR questions were converted to concept-equivalent ITs, and both sets of questions were deployed in midterm and final exams. We find that both question types provide adequate discrimination between stronger and weaker students, with CR questions discriminating slightly better than the ITs. There is some indication that any difference in discriminatory power may result from the baseline score for guessing that is inherent in MC testing. Meanwhile, an analysis of inter-rater scoring of the CR questions raises serious concerns about the reliability of the granting of partial credit when this traditional assessment technique is used in a realistic (but non optimized) setting. Furthermore, we show evidence that partial credit is granted in a valid manner in the ITs. Thus, together with consideration of the vastly reduced costs of administering IT-based examinations compared to CR-based examinations, our findings indicate that ITs are viable replacements for CR questions in formal examinations where it is desirable to both assess concept integration and to reward partial knowledge, while efficiently scoring examinations.

\end{abstract}

\maketitle
\section{Introduction}\label{sec:intro}
Constructed-response (CR) questions are a mainstay of introductory physics textbooks and examinations. Often called ``problems", these questions require the student to generate an acceptable response by demonstrating their integration of a wide and often complex set of skills and concepts. To score the question, an expert must interpret the response and gauge its level of ``correctness". Conversely, in multiple-choice (MC) testing, response options are provided within the question, with the correct answer (the keyed option) listed along with several incorrect answers (the distractors); the student's task is to select the correct answer. Because response-interpretation is not required in scoring MC items, scoring is quicker, cheaper, and more reliable \cite{Scott, Haladyna2004, wainer1993}, and these factors contribute to the increasing use of MC questions in introductory physics exams \cite{Aubrecht,Scott,Tobias}.

With proper construction, MC questions are powerful tools for the assessment of conceptual physics knowledge \cite{Aubrecht,Hestenes}, and there are examples of introductory physics final exams that consist entirely of MC questions \cite{Scott}. These tend to be in universities with large class sizes, where the procedural advantages of MC testing are weighed against any pedagogical disadvantages stemming from an exam that necessarily measures compartmentalized conceptual knowledge and calculation procedures. Conversely, MC questions are not typically used to assess the complex \textit{combination} of cognitive processes needed for solving numerical problems that integrate several concepts and procedures. Such problems involve the integration of a sequential flow of ideas---a physical and mathematical argument of sorts---that can initially seem to resist partitioning into MC items \cite{ding_2009,Ding2009}. Furthermore, the explicit solution synthesis required by CR questions gives a strong sense of transparency of student thinking that is often lacking in the MC format. For all of these reasons, the use of MC questions for formal assessments in physics education remains limited, and greater exam weight is typically placed on traditional CR questions that involve problem solving and explicit synthesis. Nonetheless, administering MC exams is considerably less time consuming and costly than administering CR exams, and the disparity of cost scales rapidly with the number of students \cite{wainer1993}. It is estimated that administering a 3-hour IT exam employing the IF-AT response system as described below costs approximately \$0.35/student (including grading and manual data entry), while an equivalent CR exam scored by a single student rater costs at least  \$7.50/student. Duplicate scoring and/or extensive training of scorers significantly increases these costs. Thus, the cost to administer a CR final exam is on the order of 20 times higher than that of an MC-based exam.

In order to marry the utility of MC with the validity of CR there is a need for new hybrid formats that will provide the procedural advantages of MC testing while maintaining the pedagogical advantages of using CR questions. The recent development of \textit{integrated testlets} (ITs) represents a significant effort to move in this direction \cite{slepkov2013}. ITs, which are described more fully below, involve the use of MC items within an answer-until-correct format, and are specifically designed to assess the cognitive and task integration involved in solving problems in physics. 

A traditional testlet comprises a group of MC items that share a common stimulus (or scenario) and test a particular topic \cite{Haladyna1992, Wainer, Sireci}. By sharing a common stimulus, the use of testlets reduces reading time and processing as compared to a set of stand-alone questions, and thus improves test reliability and knowledge coverage in a fixed-length examination \cite{Haladyna1992,wainer2007}. A reading comprehension testlet provides a classic example of a traditional testlet, with a reading passage being followed by a number of MC questions that probe the student's comprehension of ideas within the passage \cite{Sireci}.  A hallmark of traditional testlet theory is the requirement of item independence \cite{Haladyna1992,wainer2007}, which is necessary to avoid putting students in double jeopardy. That is, because students typically do not receive item-by-item feedback during MC testing, it would be unfair to include an inter-dependent set of MC questions in the test. Unlike a traditional testlet, an \textit{integrated testlet} is a set of MC items designed to assess concept integration both by using an answer-until-correct framework and by including items with varying levels of inter-dependence \cite{slepkov2013}. In an IT, one task may lead to another procedurally, and thus the knowledge of how various concepts are related can be assessed. This approach represents a markedly different way of using testlets. For example, whereas the items in traditional testlets (see for example questions 21-24 in Appendix C of Scott \textit{et al}. \cite{Scott}) can be presented in any order, the items in an integrated testlet are deliberately presented in a particular sequence.
\par
\begin{figure*}
\includegraphics[width=15 cm]{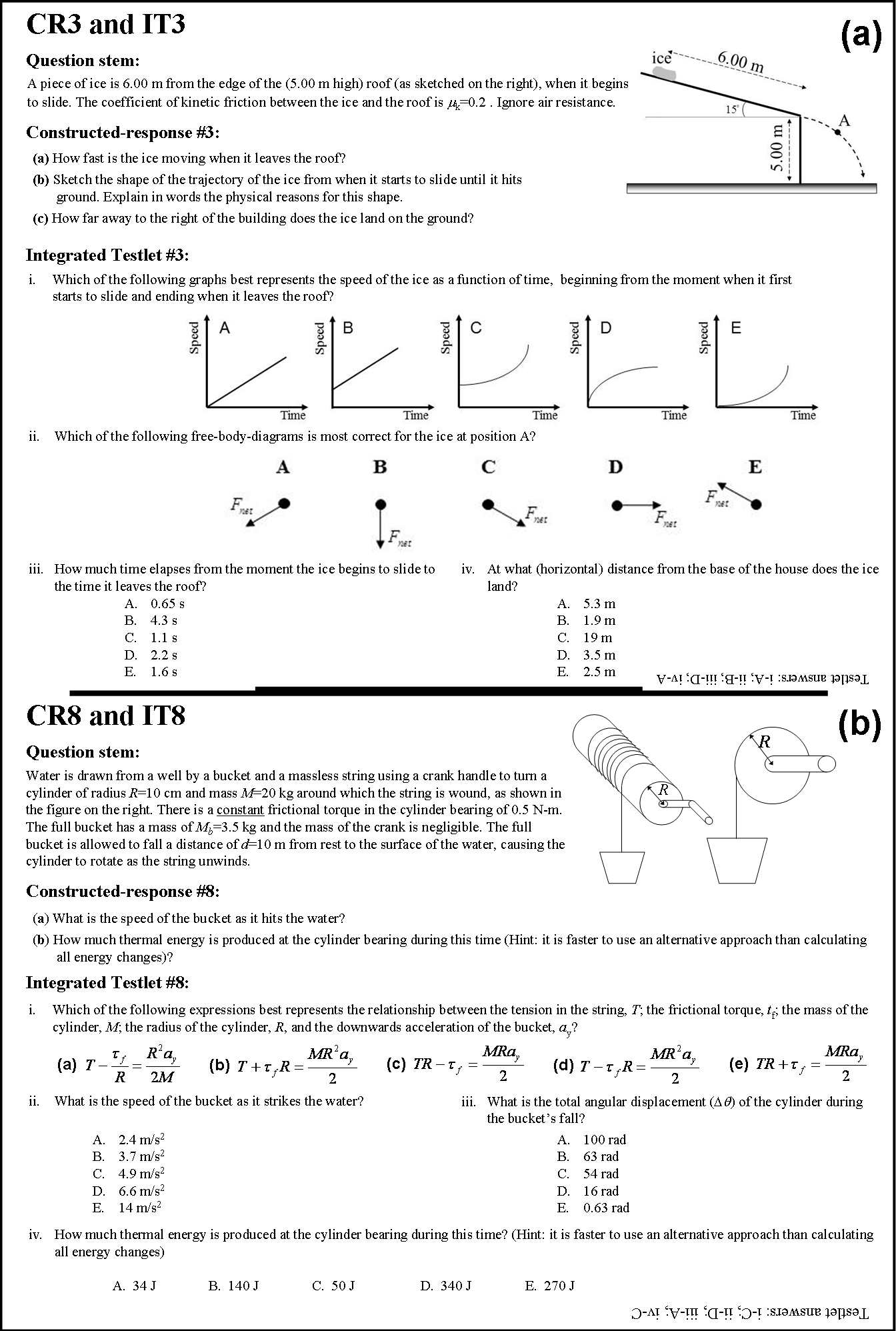}
 \caption{Examples of concept-equivalent constructed-response and integrated testlet questions. Integrated testlets comprise a set of MC items with varying levels of integration. The CR and IT questions share a common stimulus, and the final multiple-choice item in the IT is the same as the final CR sub-question. (a) CR3 and IT3 are examples from the midterm exam and cover concepts such as 2D projectile motion and kinetic friction. (b) CR8 and IT8 are examples from the final exam and cover concepts such as rotational motion, torque, and work. Note that IT8-iii exists to cue students to the most efficient means of solving IT8-iv. By contrast, such cuing is absent in CR8.}
 \label{fig:CR_and_IT_examples}
\end{figure*}

 The functional validity of ITs relies on the use of an answer-until-correct response format, wherein the responder is permitted to continue to make selections on a multiple-choice item until the correct response is both identified and can be used in subsequent related items. Certain answer-until-correct response formats, such as the Immediate Feedback Assessment Technique (IF-AT) \cite{Epstein2002-52, DiBattista2005,slepkov2013}, furthermore enable granting of partial credit within MC testing. Thus, we have designed ITs as a close proxy of traditional CR questions: both assess the complex procedural integration of concepts, and both attempt to discern contextual and nuanced knowledge by providing partial credit. Figure \ref{fig:CR_and_IT_examples} presents two examples of traditional constructed-response problems along with two integrated testlets used in the exams described below that use the same stimulus to cover the equivalent conceptual domain.

Despite the converging similarities between CR and IT, some latent differences will remain. For example, physics problems presented in CR format largely assess concept integration and synthesis, with students implicitly required to generate a tactical plan for solving the problem. In an IT, where several concepts are integrated together to build towards deeper concepts, both the order of the MC items and feedback about the correct answers to individual items suggest to students a possible procedural plan and thus remove some of the synthesis that CR assesses for (compare, for example, CR8(b) and items IT8-iii and IT8-iv in Fig. \ref{fig:CR_and_IT_examples}). 
	To establish how well ITs can act as proxies for CR questions, a direct comparison between the two is needed. The utility of ITs was recently established in a proof-of-principle study that showed that physics exams composed entirely of IF-AT-administered MC items with various levels of integration can be sufficiently valid and reliable for classroom use \cite{slepkov2013}. Here we report on a head-to-head study in which established CR questions were converted to concept-equivalent ITs, and both CR questions and ITs were simultaneously deployed in midterm and final exams of an introductory ``university physics" course. 	The purpose of this study was to address the following set of related questions: can we adequately convert traditional CR questions to ITs so as to allow for the construction of acceptable IT classroom examinations? How might we go about doing this? How fully is the divide between CR and MC bridged by such an approach, and what is gained and lost when using ITs as a replacement for CR? To address these questions we consider factors such as test statistics, testlet-level psychometrics, CR scoring procedures and inter-rater reliability issues, anonymous student surveys, and exam deployment costs

\section{Methods}
\subsection{Course structure}

A one-term introductory physics course was offered in the fall of 2012 at a primarily undergraduate Canadian university. The course instructor was one of the authors (RCS). The course is a requirement for physics, chemistry, and forensics science majors, and covers topics such as two-dimensional kinematics and mechanics, rotational motion, fluids, and heat. Course delivery followed peer-instruction and interactive-learning principles \cite{Mazur,Crouch,Meltzer}, encompassing pre-class readings followed by a just-in-time (JIT) online quiz, and in-class clicker-based conceptual tests and peer discussion. Bi-weekly laboratory sessions were alternated with bi-weekly recitation sessions at which knowledge of material covered by previous problem sets was tested with 45-minute CR quizzes, followed by tutoring of the subsequent problem-set. A two-hour midterm exam was administered during week 6 of the 12-week term, and a three-hour final exam was administered shortly after week 12; both exams consisted of a mix of CR questions and ITs. A detailed formula sheet was provided to the students at all quizzes and exams. Exams were collectively worth 50\% of a student's final grade. In total, of the 175 initial registrants, 155 students wrote the mid-term and 131 students wrote the final exam. Shortly after writing the midterm exam, students were asked to complete an anonymous online survey about their perceptions and engagement with the CR and IT question formats. Of the 155 students who wrote the midterm exam, 105 (68\%) completed the survey.

\subsection{Exam construction and scoring}
\label{subsec:Exam_construction_and_scoring}

The midterm and final exams were the experimental tools we used to directly compare CR and IT formats. However, because these also needed to be valid and fair evaluation tools within a formal course offering, particular attention was paid to balancing the questions in the experimental design. We designed two sets of complementary midterm exams (``Blue'' and ``Red'') and final exams (``Blue'' and ``Red''), where each complementary exam had an equivalent number of CR questions and ITs and covered identical course material, but swapped question formats for each topic covered. Thus, for example, the Red Midterm comprised, in order, questions CR1, IT2, IT3, and CR4, while the Blue Midterm comprised the complementary set of IT1, CR2, CR3, and IT4. Each format pair (for example, CR3 and IT3) shared a similar stimulus and covered the same material. The distribution of items among the exams is given in Table I.  Examples of complementary questions  CR3/IT3 and CR8/IT8 are shown in Fig. \ref{fig:CR_and_IT_examples}.  Each final exam included six questions; four previously unseen questions and two questions repeated verbatim from the midterm but with altered numerical values. Students were informed in advance that at least one question from their midterm was to be repeated on the final. Students were randomly assigned among the two versions of the midterm and randomized again for the final exams. Thus students were equally divided among the four possible (red/red, red/blue, blue/blue, blue/red) sequence variants.

\par Author ADS, who was not directly involved in teaching the course, designed drafts of the exams and delivered them to the instructor (RCS) two weeks before the scheduled exam time, after which both authors collaborated in editing the exams. Thus, at the time of instruction, the instructor did not know which topics would be tested in the exams. Expanded guidelines employed in constructing concept-equivalent ITs are outlined in the Appendix. In short, all of the major concepts and techniques taught in the course were listed in order of delivery, and a set of constructed-response problems taken from past exams were parsed for overlap of these concepts. Questions were then selected to give the best representation of topic and concept coverage. Rather than designing new CR questions, where possible, we chose to use final exam questions from past years to best assure some construct and content validity. Only one question (CR6/IT6) was newly created for this study because of a gap in topical coverage in recent exams.
\par Although, in principle, a testlet can include any number of MC items, each IT used in this study comprised four items, denoted, for example, IT1-i\ldots IT1-iv; each with 5 options. Each IT was constructed with its final item being stem-equivalent to the final sub-question of its matching CR question.

\par To enable the answer-until-correct MC response format needed for ITs, we used the commercially-available Immediate Feedback Assessment Technique (IF-AT)\cite{Epstein2002-52,DiBattista2005, epstein-ordering}, in similar fashion to that described previously \cite{slepkov2013}. In brief, the IF-AT provides students with immediate confirmatory or corrective feedback on each MC item as they take the test. The IF-AT response sheet consists of rows of bounded boxes, each covered with an opaque waxy coating similar to those on scratch-off lottery tickets. Each row represents the options from one MC question. For each question, there is only one keyed answer, represented by a small black star under the corresponding option box. Students provide their responses by scratching the coating off the box that represents their chosen option. If a black star appears inside the box, the student receives confirmation that the chosen option is correct, and proceeds to the next item. Conversely, if no star appears, the student immediately knows that their chosen option is incorrect, and they can then reconsider the question and continue scratching boxes until the star is revealed. It should be noted that the answer key is immutably built into the IF-AT scratch-cards and thus the MC questions presented on the exam must be constructed to match the key \cite{MultipleKeys}. This means that the IF-AT is less forgiving of minor errors in test construction than are other MC response techniques. Thus, to aid the proper construction of the tests, the mid-term and final examinations were ``test-driven'' by teaching assistants before being administered to the class.

For this comparative study, the exam scoring was designed for simplicity, with all individual MC items worth an equivalent number of marks and each IT worth the same number of marks as each CR question.  A major advantage of the IF-AT is that it enables the straightforward use of partial-credit schemes. In our MC items, for reasons outlined in section \ref{subsec:Validity_of_partial_credit}, we gave full credit (5 marks) for a correct response on the first attempt, half-credit (2.5 marks) for a correct response on the second attempt, and one-tenth credit (0.5 marks) for a correct response on the third attempt; no credit was earned for subsequent attempts. In practice, the attempt on which the correct response was attained is inferred from the number of boxes scratched and the presence of a confirmatory star within one such box. For five-option MC items, the marking scheme used can be designated as [1.0, 0.5, 0.1, 0, 0], and the expected mean item score from random guessing is 32\%.
 
To explore the typical reliability of CR scoring, we adopted two commonly-found scoring practices; that of utilizing paid student grading and that of instructor grading. CR questions on the midterm exam were scored independently by both authors, who are experienced course instructors. We did not use a common rubric, but we each scored questions in the way we considered most fair and consistent. CR questions on the final exam were scored in duplicate by two paid senior undergraduate students. In this study, they were also given detailed scoring rubrics for two of the six CR questions (CR5 and CR7) and a typical training session explaining how to score fairly and how to use these rubrics. All CR component scores reported herein represent an average of the pair of scores, otherwise known as the \textit{inter-rater} average score. 

\subsection{Exam psychometrics}
The more difficult an item, the lower the proportion of available marks that students will earn on it. A widely-used item difficulty parameter, $p$, is traditionally defined as the mean obtained item score. Typically in MC test analysis the scoring is dichotomized and $p$ is simply the proportion of the students that answer the question correctly. In our questions, where partial credit is available, a continuous or polychotomous difficulty parameter $p'$ instead represents the mean obtained question score. $p'$ ranges between 0 and 1, and its value \textbf{decreases} with question difficulty. 

At least as important as a question's difficulty is its power to discriminate between more and less knowledgeable students. Whether a question is relatively easy or difficult may be immaterial as long as the item is properly discriminating. Item discrimination is a measure of the correlation between the score on an item and the overall achievement on the test. In the case of dichotomously-scored items---such as in traditional MC items---the point biserial (PBS) correlation  value is traditionally used as a discrimination coefficient \cite{Ding_PRSTPER_2009, Haladyna2004,Ebel}. Here, however, where the availability of partial credit yields polychotomous item scores, the relevant correlation parameter is the Pearson-$r$.   

It should be noted that the correlation between the question scores and the total test scores is not between wholly independent variables \cite{guilford1954, henrysson1963}. Thus, a more pure measure of discrimination is the \textit{item-excluded discrimination parameter}, $r'$, which here is the correlation between the question score and the total test score exclusive of the question under consideration. In all cases, $r'$ is less than $r$. This distinction becomes less important as the number of questions comprising the total score increases \cite{guilford1954}, and analysis of standardized tests with $\sim100$ or more items suffers only marginally by using $r$ rather than $r'$. Given the number of questions on our exams, $r'$ is certainly the most relevant discrimination parameter for this study. While guidelines exist for interpreting the traditional item discrimination coefficient (PBS) \cite{Ding_PRSTPER_2009, doran1980}, there are currently no established guidelines for interpreting the item-excluded discrimination parameter, $r'$.

\section{Results and discussion}
\subsection{Overview}
The mean score on each version of the midterm exam was 52\%, and the means were 51\% and 50\% on the two versions of the final exam. The similarity of mean scores across versions of the exams suggests that the random divisions of the class yielded cohorts with similar overall levels of achievement, and that a comparison of achievement across exam versions is justified. 

\begin{figure}
\includegraphics[width=\columnwidth]{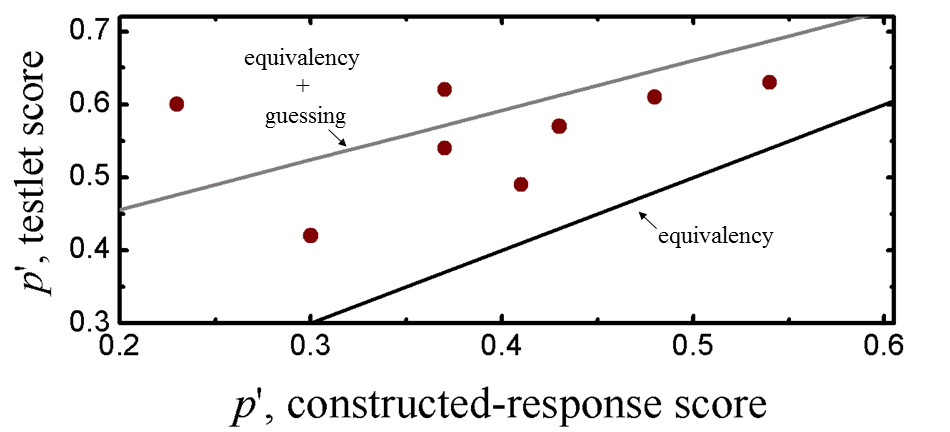}
 \caption{A comparison of item difficulty parameter, $p'$, for each matching pair of constructed response and integrated testlet questions. The solid line labeled ``equivalency'' represents the expected relationship between questions of equal difficulty. Similarly, the solid line labeled ``equivalency + guessing'' represents the case of equivalent difficulty where those students who do not know the answer choose to guess. Note that the majority of data points lie between these two lines}
 \label{fig:difficulty_correlation}
\end{figure}

There is limited data directly comparing achievement in MC and CR formats in the \textit{physics} education literature. While our sample size for the number of responders and the number of questions is somewhat limited, much can be learned from a comparison between how students engaged with concept-equivalent IT and CR questions. Table \ref{tab:summary} lists the $p'$ and $r'$ values for each IT and CR question. It is widely known that students generally obtain lower scores on CR than on MC questions, even when the stems are equivalent \cite{Reed2011, martinez1991, Barnett_Foster1996}. This finding is confirmed by our data, where (with the exception of the repeated questions) $p'$ for each IT is larger than that for the corresponding CR question. On average, the IT and CR questions yield mean $p'$ values of 0.56 and 0.39, respectively; the difference being statistically-significant, with large effect size \cite{WilcoxonDifficulty, wilcoxon1945, siegel1988}. The difference in scores between MC and CR items may be attributed to several factors: the added opportunities for guessing available in MC testing; cuing effects resulting from the presence of the correct answer among the MC options; and within our ITs, the fact that feedback provided to students using the IF-AT may enhance performance on subsequent items. Figure \ref{fig:difficulty_correlation} presents a scatter plot of $p'$ for each corresponding pair of IT and CR questions.  To estimate the plausible increase in $p'$ that arises as the result of random guessing, we can model IT questions as ones where students either know the answer \textit{a priori} or otherwise randomly guess until they find the correct answer (residual guessing). Each question has an ``inherent difficulty" assumed to be $p'_{CR}$ and if a student with an innate ability below this difficulty is presented with the question, we assume they resort to random guessing. Thus, while an ``equivalency line" would be represented by $p'_{IT}=p'_{CR}$ , an ``equivalency + guessing" line would be represented by $p'_{\text{IT}}=p'_{\text{CR}}+\bar{p}'_{\text{guess}}(1-p'_{\text{CR}})$ , where  $\bar{p}'_{\text{guess}}$ is the expected value due to guessing; 0.32 in our case. 
\par As can be seen in Fig. \ref{fig:difficulty_correlation}, all of the IT questions lie above the equivalency line, but the majority of questions lie between this line and the equivalency + guessing line. The location of any question on this figure is representative of a balance between three possible behaviors: Assuming that the inherent difficulty is indeed $p'_{CR}$, questions that are predominantly answered via the aforementioned ``know it or guess it" approach will be found scattered about the equivalency + guessing line. Questions for which partial knowledge is appropriately rewarded with partial credit will be raised above this line. Questions that contain particularly attractive ``trapping" distractors will be lowered below this line because random guessing is interrupted in favor of incorrect responses. The fact that six of eight IT/CR pairs lie between the two lines suggests that distractor trapping (see Appendix for more information) is a significant component of our carefully-constructed ITs. The two questions that are found above the top line (representing IT6/CR6 and IT7/CR7) are our best indications of questions that overall show disproportionate increase in score due to rewarded partial knowledge in the IT format. It should be pointed out, however, that from a probabilistic standpoint alone, of those points found above the top line, more will be found on the low-CR $p'$ side. Overall, this is a simplistic model, but nonetheless may provide a simple means for gauging whether a given IT is more difficult or easier than expected because of its set of distractors.

\par There tends to be a moderate-to-strong correlation between students' scores on questions in MC and CR formats \cite{Kruglak_1965,rodrigues2003,Scott}. For example, in a meta-analysis of 56 exams from various disciplines, Rodriguez found a mean correlation coefficient of 0.66 between MC and CR scores \cite{rodrigues2003}, while Kruglak  found a mean correlation of 0.59 for physics definition questions \cite{Kruglak_1965}. Our data are consistent with these findings. Specifically, we found the correlation between students' total IT and CR scores to be 0.70 and 0.83 for the two versions of the midterm and 0.69 and 0.66 for the two versions of the final exam. Thus there is evidence that, on average, our ITs operate as well or better than traditional stand-alone MC items in approximating CR questions.

\begin{table}
	\begin{threeparttable}[]
	\caption{Summary of question measures and their placement in midterm and final examinations}
\bgroup 
\def\arraystretch{1.3}
{\setlength{\tabcolsep}{0.5em}
\begin{tabular}{c c | c | c c | c c}
 \hline \hline
 \multicolumn{2}{c|}{\textbf{Question}} & \textbf{Exam} & \multicolumn{2}{c|}{$p'$ \tnote{*}} & \multicolumn{2}{c}{\textbf{$r'$ \tnote{**}}} \\
 {} & {} & {} & {\textbf{IT}} & {\textbf{CR}} & {\textbf{IT}} & {\textbf{CR}} \\
 \hline \hline
 \textbf{\textcolor{blue}{IT1}} & \textbf{\textcolor{red}{CR1}}& midterm & 0.63 & 0.54 & 0.55 & 0.64  \\ 
 \textbf{\textcolor{red}{IT2}} & \textbf{\textcolor{blue}{CR2}} & midterm& 0.61 & 0.48 & 0.69 & 0.58  \\
 \textbf{\textcolor{red}{IT3}} & \textbf{\textcolor{blue}{CR3}} & midterm & 0.57 & 0.43 & 0.60 & 0.62 \\
 \textbf{\textcolor{blue}{IT4}} & \textbf{\textcolor{red}{CR4}} & midterm & 0.54 & 0.37 & 0.21 & 0.69 \\
 \textbf{\textcolor{blue}{IT5}} & \textbf{\textcolor{red}{CR5}} & final & 0.49 & 0.41 & 0.44 & 0.70  \\
 \textbf{\textcolor{blue}{IT6}} &\textbf{\textcolor{red}{CR6}} & final & 0.60 & 0.23 & 0.51 & 0.47\\
 \textbf{\textcolor{red}{IT7}} & \textbf{\textcolor{blue}{CR7}} & final & 0.63 & 0.37 & 0.39 & 0.53 \\
 \textbf{\textcolor{red}{IT8}} &\textbf{\textcolor{blue}{CR8}} & final & 0.42 & 0.30 & 0.63 & 0.68 \\
 \hline
 \multicolumn{2}{c|}{\textbf{Mean}} & {} & 0.56 & 0.39 & 0.50 & 0.61  \\
 \multicolumn{2}{c|}{\textbf{(std. dev.)}} & {}& (0.07) & (0.10) & (0.15) & (0.08) \\
 \hline \hline
 \textbf{\textcolor{blue}{IT2$'$}} & \textbf{\textcolor{red}{CR2$'$}} & final\tnote{$\dagger$} & 0.70 & 0.70 & 0.44 & 0.61 \\
 \hline
 \textbf{\textcolor{red}{IT3$'$}} & \textbf{\textcolor{blue}{CR3$'$}} & final\tnote{$\dagger$} & 0.63 & 0.61 & 0.45 & 0.70\\
  \hline \hline
 \end{tabular}}
\egroup \label{tab:summary}
\begin{tablenotes}
\item[*] $p'$ is the ``item" difficulty parameter and is a measure of the proportion of the available score obtained by the class on a given question.
\item[**] $r'$ is the \textit{item-excluded discrimination parameter}, it is the correlation between the question score and the total test score exclusive of the question under consideration.
\item[$\dagger$] These question were repeated verbatim on the final exam from the midterm, but with changed numbers. The statistics of these questions are excluded from the combined mean and standard deviation of the other questions
\end{tablenotes}
\end{threeparttable}
\end{table}

A comparison of discriminatory properties of the IT and CR questions (see Table~\ref{tab:summary}) also seems to confirm a lesser-known relationship between MC and CR items: When the \textit{scoring} is sufficiently reliable CR questions are typically more discriminating than MC items \cite{lissitz2012, martinez1991}. Our IT and CR questions had mean discrimination scores of 0.50 and 0.61, respectively. With only 8 questions, there is insufficient statistical power to establish statistical significance in discrimination differences between CR and IT \cite{WilcoxonDiscrimination, wilcoxon1945, siegel1988}.   Nonetheless, the data suggest that our CR questions are more discriminating than ITs, with six out of eight CR questions discriminating at a greater level than their IT counterparts. This may be due in part to guessing that can take place in MC items. Alternatively, as has been identified previously \cite{martinez1991, Barnett_Foster1996}, because a significant number of students provide blank or irrelevant responses to CR questions, the effective scoring range for CR questions is larger than for ITs. For example, in our study, a score of zero was awarded on an IT only twice out of $\approx 700$ scored ITs, whereas this occurred 21 times on the same number of CR questions. However, a score of 100\% was awarded in 9\% of all instances of both IT and CR questions.  Thus, some loss in discriminatory power may be expected when replacing CR questions with ITs. 
 
While somewhat lower than that for CR, the mean IT item-excluded discrimination parameter, $r'$, of 0.50 compares favourably with MC questions typically found on classroom exams. For example, DiBattista \& Kurzawa examined 1200 MC items on 16 university tests in a variety of disciplines and found the mean (non-item-excluded) discrimination coefficient to be only 0.25 \cite{DiBattista2011}. Although we feel that the integrated testlet (as opposed to an individual MC item) is the most appropriate unit of measurement, testlet-level psychometric analysis is relatively uncommon \cite{testlet-level, Wainer}, and, in fact, combining multiple MC items into a testlet increases the discrimination parameter of the testlet over that of the average of the individual MC items comprising the testlet \cite{testlet_boost}. Thus, for comparison purposes only, we report our mean non-item-excluded discrimination coefficient for the entire set of 40 individual MC items in this study to be an impressive $\bar{r}=0.45$ \cite{average-discrimination,Ding_PRSTPER_2009}. Subsequent analysis of our items discounts a link between item interdependence (through item position within testlets) and item discrimination. Thus, the high discrimination values may arise simply due to the care with which all items were written.

\subsection{Test length, composition, and reliability}

Test reliability is a measure of how consistently and error-free a test measures a particular construct \cite{Ebel}. A set of test scores that contain a large amount of random measurement error suggests that a repeat administration of the test, or the administration of a suitable equivalent test, could lead to different outcomes. A commonly used measure of test reliability is Cronbach's $\alpha$, which provides a measure of internal consistency \cite{Webb2006, Ebel}. The theoretical value of alpha ranges from zero to one, with higher values indicating better test reliability. For low-stakes, classroom, exams an $\alpha > 0.70$ is desirable \cite{Ding_PRSTPER_2009, doran1980}. Both versions of our final exam yielded $\alpha =0.79$, indicating that a mixed-format final exam with three CR and three IT questions can provide satisfactory reliability. In principle, we could assess individually how the IT and the CR portions differentially contribute to the exam reliability, and thus discern which question format is more reliable. However, there is insufficient statistical power to make such a determination in the current study.  As a guiding principle, the reliability of a test scales with the number of items (via, for example, the Spearman-Brown prediction formula \cite{guilford1954_SB}). Thus, one minor advantage of using ITs over CR may be in the ability to include more items in a fixed-length exam, as preliminary indications suggest that students spend less time completing an IT than they do completing its complementary CR question \cite{tutorials}. The anonymous student surveys support this notion. When asked: ``Regardless of which type of question (the testlets or short-answer questions) you found more difficult, which did you spend the most time on?", 48\% of students indicated that CR questions took longer to complete, 22\% indicated they spent the same amount of time on CR and IT questions, and 27\% indicated that ITs took longer to complete. Because test reliability scales with the number of questions, then in order to create optimally-reliable IT-only exams, one could add more questions while maintaining test duration. This does not, however, assure that an exam solely comprising ITs will be more reliable than a mixed-format exam containing both IT and CR components. Furthermore, only sufficient gains in question completion times would motivate such an approach. In the past, analysis of which question format, CR or MC, is inherently more reliable has largely proven inconclusive. While some studies find that MC is more reliable, others find the converse \cite{wainer1993, rodrigues2003}. We suspect that the relative reliability between MC and CR depends strongly on the balance between the strength of MC item writing and on the consistency of CR scoring \cite{wainer1993}.

\subsection{Inter-rater reliability}

The manner by which each question contributes to test reliability hinges on any randomness in the scoring of each component. In multiple-choice testing, the contribution of guessing to the total score dominates the discussion of the (un)reliability of the method. On the other hand, while a constructed response may be a more faithful reflection of the responder's state of knowledge, the \textit{interpretation} of that response can be highly subjective, thus diminishing the reliability of the question score. This subjectivity in scoring is inherent to the CR format, but is rarely mentioned when comparing the attributes of CR and MC formats in classroom examinations. As part of our formal comparison of the relationship between CR and IT formats we have assessed the effects of inter-rater reliability on CR score reliability. As mentioned above, the CR components of the midterm exam were scored in duplicate by two professors, without a shared rubric, while the CR components of the final exams were scored in duplicate by two paid student graders, who shared a formal rubric for two of their six assigned questions. Inter-rater scoring data is presented in Table~\ref{tab:interrater}.

Traditionally, a correlation coefficient between the scores of two raters is used as a measure of inter-rater reliability \cite{Kruglak_1965, Barnett_Foster1996, Hancock1994}. The correlation coefficient for the inter-rater scoring of every CR question in our study ranges from $r=0.79$ to $0.95$. On first inspection, it may seem that these high correlations imply strong inter-rater reliability. In the only other similar comparison between MC and CR components on a physics classroom exam we have found, Kruglak reports a similar range of correlations between course instructors scoring in duplicate \cite{Kruglak_1965}. The strong inter-rater correlation does imply that in general raters \textit{rank} students' question scores consistently. However, a closer inspection of student scores suggests a larger amount of both systematic and random variability between raters, as shown in Table \ref{tab:interrater}. We find, for example, that the mean difference in question scores ranges from -6 to +15 percentage points. Such a sizable mean difference is an indication of inter-rater \textit{bias}; wherein one rater systematically scores an item higher than the other rater does. Whereas Kruglak found bias on the order of 4 to 8 percentage points  for questions related to physics definitions, we find bias between raters as high as 15 percentage points for paid student scorers for these more traditional physics ``problems". Such bias does not affect the inter-rater correlation measures, as they are systematic and may not affect the ranking of total scores. On the other hand, the standard deviation in the differences between the scores makes more apparent the measure of random error in scoring. This value ranges from 11 to 17 percentage points. This clearly represents a latent ``unreliability" in CR scoring that is not often addressed in the literature. This effect is not merely tied to the common (if non-optimal) practice of using non-expert scorers. When the final exam scorers were instructed to use a detailed rubric, the inter-rater reliability did not improve. Thus, while there is an element of unreliability to what responses mean in MC items, there is also an element of unreliability in the interpretations of responses in CR questions. Classroom tests rarely address this limitation of CR testing. Conversely, scoring ``high-stakes tests" often requires great efforts and cost to minimize inter-rater variability \cite{Maryland}, which in large part has motivated the shift towards pure MC testing in standardized tests \cite{wainer1993}.

\begin{table*}
	\begin{threeparttable}[b]
	   \caption{Inter-rater (IR)\tnote{+} reliability for constructed-response questions}
		\def\arraystretch{1.3}
		{\setlength{\tabcolsep}{0.8em}
		\begin{tabular}{c p{0.8cm} | p{1.7 cm} p{1.7 cm} p{1.7 cm} p{1.8 cm}}
		 \hline \hline
		\textbf{Question} & \textbf{} & \textbf{IR \newline correlation, \emph{r}} & \textbf{IR \newline mean \newline difference} & \textbf{IR \newline difference \newline standard \newline deviation \tnote{$\ddagger$}} & \textbf{ IR \newline difference \newline extrema \newline pos./neg.} \\
 		\hline \hline
 		\multirow{4}{*}{MIDTERM \tnote{*}} & \textcolor{red}{CR1} & 0.92 & -6\%\tnote{$\dagger$} & 13\% & 						+55\%/-20\%\\
 		{} & \textcolor{blue}{CR2} & 0.95 & +3\%\tnote{$\dagger$} & 11\% & +33\%/-25\%\\
 		{} & \textcolor{blue}{CR3} & 0.91 & +6\%\tnote{$\dagger$} & 11\% & +35\%/-25\%\\
 		{} & \textcolor{red}{CR4} & 0.87 & -1\% & 17\% & +48\%/-33\%\\
 		\hline
		 \multirow{6}{*}{FINAL \tnote{**}} & \textcolor{red}{CR2$'$}\tnote{$\blacklozenge$} & 0.92 & 		+11\%\tnote{$\dagger$} & 15\% & +45\%/-13\%\\
 		{} & \textcolor{blue}{CR3$'$}\tnote{$\blacklozenge$} & 0.87 & +3\% & 14\% & +55\%/-38\%\\
		 {} & \textcolor{red}{CR5}\tnote{$\lozenge$} & 0.90 & +15\%\tnote{$\dagger$} & 16\% & +65\%/-5\%\\
		 {} & \textcolor{red}{CR6} & 0.79 & +15\%\tnote{$\dagger$} & 17\% & +73\%/-13\%\\
		 {} & \textcolor{blue}{CR7}\tnote{$\lozenge$}& 0.88 & +9\%\tnote{$\dagger$} & 11\% & +65\%/-30\%\\
		 {} & \textcolor{blue}{CR8} & 0.88 & +1\% & 17\% & +33\%/-45\%\\
		 \hline \hline
		 \end{tabular}}\label{tab:interrater}
	\begin{tablenotes}
	\item[+] All IR differences refer (arbitrarily but consistently) to ``scorer 1"-``scorer 2".
	\item[$\dagger$] A statistically-significant measure of inter-rater bias, as determined by a paired-item t-test ($p<0.05$).
	\item[$\ddagger$] a measure of average random error in item scoring
	\item[*] midterm exam scorers are professors
	\item[**] final exam scorers were hired senior undergraduates
	\item[$\blacklozenge$] Repeats on the final exam of midterm CR2 and CR3
	\item[$\lozenge$] Items for which scorers were provided with a detailed scoring rubric
	\end{tablenotes}
	\end{threeparttable}

\end{table*}

\subsection{Validity of partial credit in IT scoring}
\label{subsec:Validity_of_partial_credit}

A key difference between CR and MC testing has traditionally been the means of question scoring. While MC questions are almost invariably scored dichotomously, assessment of partial credit has been a mainstay of traditional physics CR questions. The unavailability of partial credit as a means of rewarding substantial (if incomplete) knowledge is largely seen as a major drawback of traditional MC testing, as it severely limits the assessment of complex knowledge integration. The answer-until-correct framework of IT usage allows for the granting of partial credit for questions in which students initially respond incorrectly, but ultimately respond correctly on subsequent attempts. The validity of such an approach depends on whether the partial credit is being assessed in a discriminating manner; whether it reliably represents some measure of partial knowledge. A previous study of IF-AT-administered exams that utilized a heterogeneous mix of stand-alone items and integrated testlets found that such discrimination is possible \cite{slepkov2013}. In this current study, as in the former, there is an inverse correlation between the amount of partial credit granted to any given student and their exam score. This is mostly due to opportunity; the top scorers are more likely to get full credit on any question and thus have fewer opportunities to earn partial credit. Nonetheless, the granting of partial credit proves discriminating. To demonstrate this, we consider the likelihood that a student earns the \textit{available} partial credit. Only in cases when a first response is incorrect does a student have the opportunity to earn partial credit. When partial credit is used in a discriminating manner, we expect top students to earn a higher proportion of their available partial credit as compared to the students at the bottom. As a good means of measuring the discrimination afforded by partial credit, we would ideally use the correlation between the total IT score for each student, scored dichotomously, and the percentage of available partial credit converted. However, the strong inverse relationship between the dichotomously-scored total and the opportunity for earning partial credit means that many of the top students do not have much opportunity to earn partial credit, thus making such analysis less robust. Instead, we use a median-split analysis \cite{median-split,Maccallum2002}, which relies on comparisons between the (dichotomously-scored) top and bottom 50\textsuperscript{th} percentile groups. As shown in Table \ref{tab:partial-credit}, in all exams the top students converted a higher percentage of available partial credit, as compared to the bottom group. A t-test confirms that for three of four exams this difference is significant at the $p<0.05$ level. We also note that with [1, 0.5, 0.1, 0, 0] scoring, blind guessing for partial credit is expected to yield a conversion rate of 30\% of available partial credit. As a cohort, the top scorers obtain partial credit at a much higher rate than this, converting on average 56\% of available partial credit, and thus are not likely randomly guessing, but instead are demonstrating partial (or corrected) knowledge of the answers. Furthermore, in all exams, the bottom half of the class also earned partial credit at a higher rate than expected by random guessing.

\begin{table}
\caption{Analysis of partial credit granted within ITs}
\bgroup
\def\arraystretch{1.5}
{\setlength{\tabcolsep}{0.5em}
\begin{tabular}{l | p{1.4 cm} p{2.2 cm} p{1.5cm} }
 \hline \hline
 \textbf{Exam} & \textbf{\# \newline students} & \textbf{available \newline partial-credit \newline converted by \newline top/bottom \newline half of class} &  \textbf{t-test, \newline $p$-value}
\\
 \hline \hline
 \textbf{\textcolor{blue}{Midterm-B}} & 82 & 50\%/35\%  & 0.001 \\
 \hline
 \textbf{\textcolor{red}{Midterm-R}} & 73 & 64\%/44\%  & 0.001 \\
 \hline
 \textbf{\textcolor{blue}{Final-B}} & 63  & 56\%/47\% & 0.12 \\
 \hline
 \textbf{\textcolor{red}{Final-R}} & 68 & 52\%/41\% & 0.025 \\
  \hline \hline
 \end{tabular}}
\egroup
\label{tab:partial-credit}
\end{table}

The choice of marking scheme, and therefore the proportion of partial credit granted, will influence the overall mean test score and may influence the discriminatory power of the granting of partial credit \cite{DiBattista2009}. In this study we used a [1, 0.5, 0.1, 0, 0] scoring scheme, but could have used any number of alternate schemes. There are currently no well-established or research-derived guidelines for the best scoring schemes in examinations that utilize the IF-AT \cite{DiBattista2009}. Over time, we have converged to the current scheme in an attempt to balance a desire to keep the expectation value for guessing sufficiently low as to make passing of the test statistically unlikely with guessing alone, with a desire to prolong students' intellectual engagement with the questions via partial credit incentives. While these two considerations are largely distinct, the choice of scoring scheme must balance the two. From consultation with students we have discerned that offering students an opportunity to ``pass" the question by giving them 50\% or more on a second attempt is a significant incentive to remain engaged with a question beyond an initial incorrect response. We chose to offer precisely 50\% to take advantage of this effect while moderating the overall test score. We then offer 10\% on a subsequent correct response in an attempt to keep more students engaged further in the question, again without substantially increasing the overall test score. We have not utilized a scheme that rewards students past a third incorrect attempt because we suspect that at this point students are either likely to revert to random guessing, or that whatever partial knowledge they use to answer the question at this point will no longer be discriminating. Whether these considerations are strictly justified is an avenue for future research \cite{DiBattista2009}. Regardless of \textit{a priori} motivations for using this scheme, a \textit{post hoc} analysis of alternate hypothetical scoring schemes proves illuminating. Table \ref{tab:Scoring Rubrics} lists various plausible IF-AT scoring schemes, and compares their effects on the average obtained testlet scores ($\bar{p'}$) and discrimination measures ($\bar{r'}$) for the testlets deployed in the final exams. None of the alternate schemes under consideration include any partial credit beyond the third response because in the actual exam students did not have this option, and thus their fourth/fifth responses are indiscernible. As presented in the table, we considered schemes that range from dichotomous (all or nothing), through a ``harsh" scheme that grants a modicum of partial credit for a correct second response, to a ``generous" scheme that grants more partial credit for second and third correct responses than in our as-given [1,0.5,0.1,0,0] scheme. The choice of partial-credit scheme directly affects the average test score. For our exams, the difference between the dichotomous scoring and most generous scoring schemes is 20 percentage points, with an MC component score of 45\% for the former and 65\% for the latter scheme. As given, the MC component score is 58\%. All other schemes considered are much closer to the as-given scheme than to either of the extreme cases. On the other hand, inspection of the effect of the scoring scheme on the discriminatory power of the questions reveals this measure to be quite robust, and $\bar{r'}$ only ranges from 0.45 to 0.48. It is noteworthy that the most generous scoring scheme shows the lowest discriminatory power, while our standard scheme proves as or more discriminating that the others. It should also be noted that the fact that the as-given test proves as discriminating as the (\textit{post hoc}) dichotomously-scored test is evidence that partial credit is \textit{at least} as discriminating as the first-response credit. Overall, it appears that all of the plausible schemes considered are viable from a discriminatory standpoint, and thus the main considerations for their adoption are the targeted exam score and student reception.

\begin{table}
\caption{Effects on average testlet score and discrimination of various hypothetical MC item scoring schemes. All six testlets used on red and blue final exams are considered. ``(change)" denotes deviation from that obtained using the actual, as given, scoring scheme}
\bgroup
\def\arraystretch{1.5}
{\setlength{\tabcolsep}{0.35em}
\begin{tabular}{l | c c c}
 \hline \hline
 \textbf{Scheme} & \textbf{$\bar{p'}$ (change)} &  \textbf{$\bar{r'}$ (change)} & \textbf{notes}
\\
 \hline \hline
 [1,0.5,0.1,0,0] & 0.58 (N/A) & 0.48 (N/A) & as-given \\
 \hline
 [1,0.5,0,0,0] & 0.56 (-0.02) & 0.48 (0.00)  & ``Two-strikes" \\
 \hline
  [1,0,0,0,0] & 0.45 (-0.13) & 0.47 (-0.01)  & Dichotomous \\
 \hline
  [1,0.6,0.2,0,0] & 0.59 (+0.01) & 0.47 (-0.01)  &  \\
 \hline
  [1,0.6,0,0,0] & 0.59 (+0.01) & 0.47 (-0.01)  &  \\
 \hline
  [1,0.4,0.2,0,0] & 0.57 (-0.01) & 0.48 (0.00)  &  \\
 \hline
  [1,0.7,0.3,0,0] & 0.65 (+0.07) & 0.45 (-0.03)  & ``Generous" \\
 \hline
 [1,0.3,0,0,0] & 0.52 (-0.06) & 0.48 (-0.00)  & ``Harsh" \\
 \hline
 \end{tabular}}
\egroup
\label{tab:Scoring Rubrics}
\end{table}

\subsection{Correlational evidence of similarity between the operation of CR and IT formats}

The comparison of how and what CR and MC formats measure in test takers has been an active area of research \cite{rodrigues2003, Scott, martinez1991, Lin_WeightedMCscoring_2012, Hancock1994, Bennett1991}. While there is a strong sense from some, including physicists, that the two formats measure fundamentally different things, much research has concluded that there is little evidence to support this notion \cite{wainer1993,rodrigues2003}. One of the main research questions addressed by this study is to gauge whether, by the use of ITs, MC can be made more like CR from a content and cognitive domain standpoint. Thus, it is imperative to get a sense for whether ITs and CR questions act in fundamentally distinct ways, or whether they are largely acting similarly but with slightly different performance measures. To address this issue we construct a correlation matrix that describes the correlation between each item score on a given exam to every other item on that exam.

\begin{table}
\caption{Correlation table for scores of CR and IT questions in the final exams. The upper triangle lists the intra-test question correlation coefficients for the ``red'' final exam, and the lower triangle lists those for the ``blue'' final exam. For example, for the upper triangle the row and column labels ``2\superscript{nd} IT'' each refer to IT3, and for the bottom triangle these refer to IT4, as these are the second IT within each respective exam.}
\bgroup
\def\arraystretch{1.0}
{\setlength{\tabcolsep}{0.0em}
\begin{tabular}{p{1.2 cm} | p{0.7 cm} p{0.7 cm} p{0.7 cm} p{0.7 cm} p{0.7 cm} p{0.7 cm} }
 \hline \hline
  & \textbf{ 1\superscript{st} CR} & \textbf{ 2\superscript{nd} CR} & \textbf{ 3\superscript{rd} CR} & \textbf{ 1\superscript{st} IT} & \textbf{ 2\superscript{nd} IT} & \textbf{ 3\superscript{rd} IT}\\
 \hline
 \textbf{1\superscript{st} CR}  & 1 & \cellcolor{red!60}0.62 & \cellcolor{red!60}0.41 &\cellcolor{red!60} 0.39 &\cellcolor{red!60} 0.17 &\cellcolor{red!60} 0.48  \\
 \textbf{2\superscript{nd} CR}  & \cellcolor{blue!60}0.47 & 1 & \cellcolor{red!60}0.39 & \cellcolor{red!60}0.51 & \cellcolor{red!60}0.29 & \cellcolor{red!60}0.54  \\
 \textbf{3\superscript{rd} CR}  & \cellcolor{blue!60}0.56 & \cellcolor{blue!60}0.51 & 1 & \cellcolor{red!60}0.15 & \cellcolor{red!60}0.35 & \cellcolor{red!60}0.40  \\
 \textbf{1\superscript{st} IT}  & \cellcolor{blue!60}0.50 & \cellcolor{blue!60}0.37 & \cellcolor{blue!60}0.41 & 1 & \cellcolor{red!60}0.23 & \cellcolor{red!60}0.29  \\
 \textbf{2\superscript{nd} IT}  & \cellcolor{blue!60}0.42 & \cellcolor{blue!60}0.27 & \cellcolor{blue!60}0.39 & \cellcolor{blue!60}0.14 & 1 & \cellcolor{red!60}0.46  \\
 \textbf{3\superscript{rd} IT}  & \cellcolor{blue!60}0.50 & \cellcolor{blue!60}0.28 & \cellcolor{blue!60}0.51 & \cellcolor{blue!60}0.15 & \cellcolor{blue!60}0.40& 1  \\
  \hline \hline
 \end{tabular}}
\egroup
\label{tab:correlation_table}
\end{table}

Table \ref{tab:correlation_table} presents such a matrix for the two final exams. Overall, all items correlate positively with all other items on the exam, consistent with our discrimination analysis that identified that all CR and IT questions had positive discriminations. The correlations range from 0.14 to 0.62. Comparing the median CR-CR, CR-IT, and IT-IT correlation is highly suggestive that IT and CR items do not behave in fundamentally separate ways. For the red exam, the median item correlations are 0.41, 0.39, and 0.29, respectively for CR-CR, CR-IT, and IT-IT. Likewise, for the blue exam the values are 0.51, 0.41, and 0.15. Thus, it is clear that while (on average) a CR question behaves most similarly to other CR questions, so too does the average IT. Scores on ITs correlate more closely to those on CR questions than they do to other ITs. This suggests that while the CR format is perhaps measuring what we care about better than does the IT format, the two formats do not measure fundamentally different things. Were these two formats behaving in fundamentally different ways---i.e., accessing different testing ``factors"---we would expect the IT-IT median correlations to be higher than the IT-CR median correlations.  Factor analysis would be a more direct and robust way to gauge this, but it would also require a much larger study. Thus, while CR and IT questions do not perform to the same level of discrimination, they do not seem to perform distinct measurement tasks, and are hence similar in how they measure the desired construct.

\subsection{Limitations of the study and future directions}

This study answers a number of key questions concerning whether or not IT structures can replace traditional CR questions on formal exams. Our study involved $\approx 150$ students, which is  triple the size of the previous pilot study \cite{slepkov2013} and presents for the first time a direct comparison of concept-equivalent CR and IT questions. Nonetheless, many of our results are only suggestive of the differences and similarities between IT and CR. Additional head-to-head testing between CR questions and concept-equivalent ITs is needed to better establish statistical significance between their discriminatory powers. This need is independent of the number of students in the study, and can only be met by deploying and analyzing more CR/IT pairs.

\par A key difference between CR and IT questions has so far been left unexamined: The procedural cuing implicit in the question order within an IT reduces the testing of solution synthesis that is such a powerful aspect of CR. We have not investigated this nuanced question, which will be addressed in future work. Likewise, the formative assessment nature of ITs has only been hinted at as a key attribute of the tool, \cite{DiBattista2009,Dihoff,Brosvic2005} and establishing the extent to which ITs can prove formative will also be addressed in the future.
	This study aims to compare head-to-head CR and IT formats in an effort to bridge the divide between CR and MC tests. However, no attempt has been made here to compare IT and stand-alone MC questions. This is largely due to our presumption that due to the limited cognitive complexity assessed by typical MC tests, they do not have the \textit{construct validity} we are looking for in a CR physics test. MC tests may \textit{reliably} test something that we are only partially interested in testing. With this study we indicate that a concept-equivalent IT test can measure something much closer to what we want, but possibly with reduced reliability.
There has been an ongoing desire to better establish the relationship between CR and MC testing formats by direct comparisons of stem-equivalent questions \cite{rodrigues2003}. Such comparisons, where the only differences between a CR and MC question lies in the availability of response options within the MC item, are the most direct means of measuring differences due purely to the question format, rather than to content or contextual differences. While some of our items are stem-equivalent with CR sub-questions (for example IT8-ii/CR8(a) and IT8-iv/CR8(b); as shown in Fig. \ref{fig:CR_and_IT_examples}), we cannot directly use our data for a valid stem-equivalent comparison for several reasons: First, not all of our CR sub-questions have a strictly stem-equivalent MC item match (for example IT3/CR3; as shown in Fig. \ref{fig:CR_and_IT_examples}). Second, even when sub-questions are stem-equivalent with a testlet item, there are contextual differences between the items that make such comparisons difficult. For example, the lack of immediate feedback typically leads to sub-questions within a CR question that appear more difficult because of aforementioned-multiple jeopardy issues. A complete comparison of stem-equivalent CR and IT questions would at the least require either a means for providing immediate feedback in the CR portion of the exams or the introduction of ``dummy values'' in the CR sub-question stems, in addition to the strict construction of all items as verbatim stem-equivalent. This study, on the other hand, compares \textit{concept-equivalent} questions; where the same concept and procedure domains are tested.  Thus, this comparison is meant as a more valid comparison between question format than one would get by comparing an arbitrary set of IT and CR questions, but nonetheless presents an incomplete picture of effects of the question format on its discrimination, and the test reliability.
 
\section{Summary and conclusions}

There is a dearth of formal comparisons between multiple-choice and constructed-response question formats in science education. The recent development of ``integrated testlets"---a group of inter-dependent MC items that share a stem and which are administered with an answer-until-correct response protocol---has been described as a possible replacement for CR format questions in large classroom assessments \cite{slepkov2013}. In this study we directly compare the administration of concept-equivalent CR and IT questions in formal classroom exams. We find that scores on ITs are higher than those of equivalent CR questions, but the difference is small and generally within the range accounted for by some of the opportunities for guessing inherent to multiple-choice formats. We find that both CR and IT questions can be highly discriminating and reliable in their assessment of introductory physics knowledge, with the CR format appearing marginally better at both of these measures. A 3-hour mixed-format exam proves to be more than sufficiently reliable for a classroom exam. While a pure CR exam may prove marginally more reliable than a pure IT exam with the same number of questions, because ITs take less time to complete, more questions may be employed to increase the test reliability. A comparison of inter-rater reliability of two individuals scoring CR exams in duplicate reveals that while the score correlations between them is high, there is large latent random and systematic variability in scores. This kind of data are rare in the literature, and raise important questions of reliability and validity when using multi-step CR questions as primary assessment tools. The answer-until-correct response format used for administering ITs allows for straightforward granting of partial credit within the auspices of a multiple-choice test, and we provide evidence that the granting of partial credit is accomplished in a discriminating manner. The ability to assess partial knowledge with IT structures goes a long way towards bridging the divide between CR and MC formats. Finally, an analysis of the correlation between CR and IT scores dispels notions that ITs and CR questions measure distinctly different constructs, but rather suggests that while CR questions are more reliable than IT questions, both types of questions largely measure the same thing. On average IT scores correlate more closely to other CR scores than to other IT scores.
	
\par Beyond any suggestions that for a given exam duration the CR format may prove both more reliable, discriminating, and is \textit{a priori} of higher construct validity, one important comparison remains; that of cost, which is on the order of 20 fold higher for CR than IT exams. We have shown that ITs approximate CR questions and yield comparable measures of reliability, validity, and discrimination, and thus, in light of the disparity in costs, ITs are a viable proxy for CR questions for formal assessments in large classes.

\begin{acknowledgments}
We thank David DiBattista of Brock University for guidance and advice on the experimental design, and for extensive manuscript editing. We thank Angie Best of Trent's Instructional Development Centre for her support. This work was funded by a Trent University Learning Innovation Grant. We acknowledge Greg Hodgson for assistance in testlet construction and experimental design. We also thank Dr. Bill Atkinson and  Dr. Rachel Wortis for useful discussion and encouragement.  
\end{acknowledgments}

\section*{APPENDIX: Guidelines for integrated testlet design}

To create concept-equivalent IT and CR pairs we started with a set of CR problems taken from past exams, deconstructed the concepts and procedures needed to solve the problem, weighted the importance and difficulty of each part much as one would when constructing a scoring rubric, and created four multiple-choice items that addressed one or two specific conceptual or numerical steps in the solution. Ultimately, the choice of how many items comprise a testlet and which steps in the solution we wish to include in the testlet is based on time constraints and on a targeted difficulty level. 

\begin{figure}
\includegraphics[width=\columnwidth]{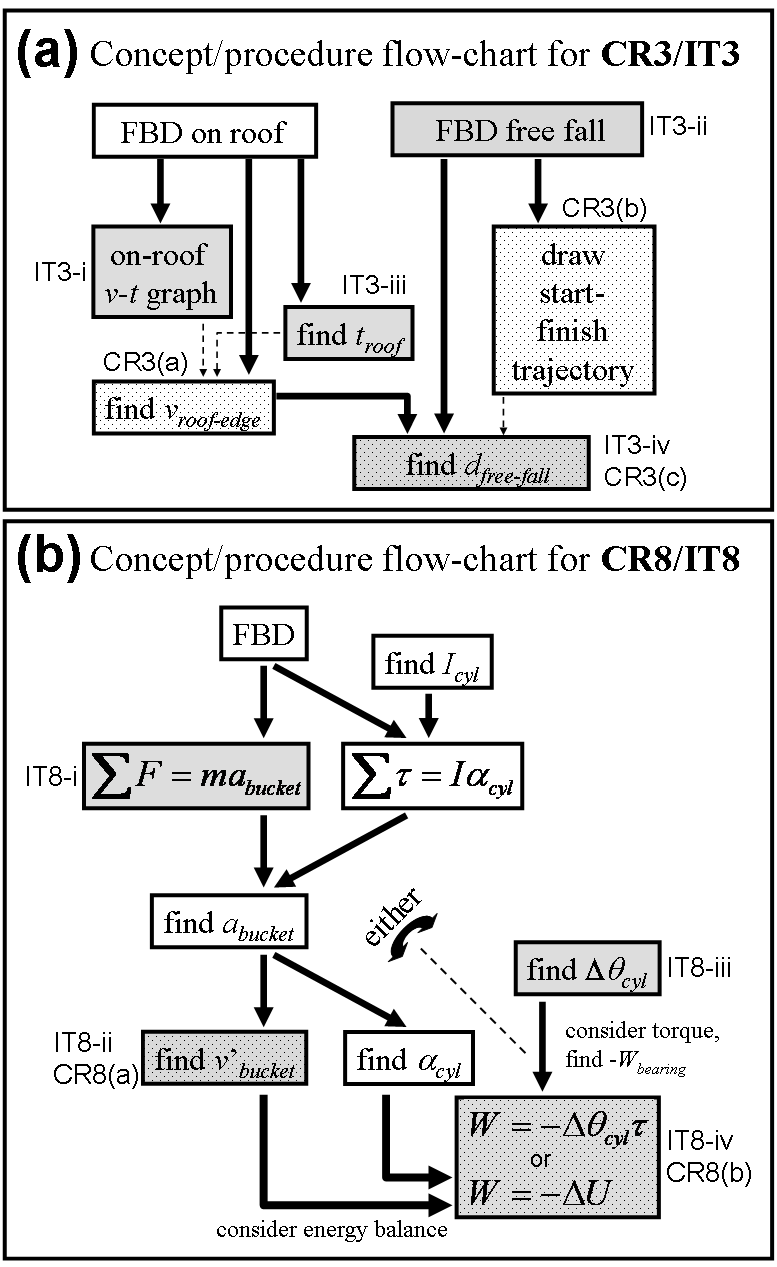}
 \caption{A conceptual and procedural map of the two concept-equivalent exam structures shown in Figure~\ref{fig:CR_and_IT_examples}. The various sub-items in the constructed-response are labeled and highlighted by dotted stipling. Individual items within a testlet are labeled and highlighted with gray shading.  (a) Constructed response question 3(a-c) (CR3) and a 4-item testlet (IT3). FBD=free-body diagram; $d_{free-fall}$=horizontal ice landing distance from roof edge. (b) Constructed response questions 8(a,b) and 4-item testlet 8 (IT8). cyl=cylinder; $v'_{bucket}$=bucket's speed at ground height. Arrows indicate which concepts and parameters are needed for developing other concepts and parameters. Two alternate conceptual approaches to final step are indicated. Unlike in the CR question, the integrated-testlet cues and builds scaffolding for an easier approach to the final question.}
 \label{fig:testlet_maps}
\end{figure}

	Figure \ref{fig:testlet_maps} provides a visual map representing this procedure for CR3/IT3 and CR8/IT8, which are reproduced in Fig. \ref{fig:CR_and_IT_examples}. We have identified seven non-trivial ``elements" in the solution of CR3, and nine in the solution of CR8. In each solution map, we chose four key elements to include as individual MC items, as indicated in the figure. All CR questions used in this study had at least two sub-questions, with the solution to the later ones often depending on previous answers. For example, CR3(a) and CR3(b) are independent, but CR3(c) is weakly dependent on CR3(b) and strongly dependent on CR3(a), as depicted in Fig. \ref{fig:testlet_maps}. In creating a 4-item integrated testlet from this question we deemed that CR3(c) is the intended destination of the problem, and thus include it as the final testlet item, denoted IT3-iv. However, the testing of intermediate steps does not necessarily have to follow that of the CR question, and in IT3 we chose three different intermediate elements to test. In this sequence, we do not expect the question to be particularly difficult, and thus the intermediate steps are dispersed and not strongly integrated. It is expected that when the items are strongly integrated and where the final item depends strongly on a particular preceding step, that including this step as an item makes the question easier. This aspect of the answer-until-correct approach mirrors that of Ding's ``conceptual scaffolding" question sequences \cite{ding_2009,Ding2009}, where CR questions that involve the particular integration of multiple disparate concepts are preceded by short conceptual MC items that implicitly cue the students to consider those concepts. Thus, Ding's question sequences also utilize an integrated question formalism but without the implementation of immediate feedback or partial credit. We too rely on items within a given testlet to act as scaffolding for other items in the testlet. 
	
	The issue of how distractors are created is also related to intra-question scaffolding and discrimination. There are several ways in which distractors can be created: For numerical answers, distractors can be quasi-randomly chosen values; they can represent answers obtainable via rational missteps (i.e. identifiable mistakes); and they can be responses that are selected because of their relationship to other distractors. The choice of approach taken for creating any given distractor lies in the assessment objectives of any given question. For example, if a key concept being tested for is the quadratic (as opposed to linear) relationship between two variables, including a distractor that results from a linear analysis may be warranted, as it should aid in discriminating for the key concept. On the other hand, neglecting to ``trap" for such linearity by omitting such a distractor is also tantamount to creating scaffolding within a question. Finally, creating a distractor that results from neglecting to implement a trivial procedure (such as doubling a result) may simply represent a non-discriminating ``trap" to be avoided. Thus, when choosing discriminators, it is important to also consider the assessment objectives and concept maps underpinning any given question.  

The concept of scaffolding is of prime importance to our philosophy of integrated testlets: Ultimately we wish to use MC structures to test how well a student can climb to the apex of a ``mountain of knowledge". Typically in a multiple-choice exam, we are relegated to surveying the perimeter of this mountain. With CR questions we often ask the student to climb the mountain, but when a student falters early in the process, they have few tools to assist their climb, and thus we cannot adequately test subsequent progress. With an answer-until-correct integrated testlet we can assess the student's climb from the base to the apex, providing the needed scaffolding as they ascend. Students who do not need the scaffolding get all questions correct on the first try. Some students, however, need help in particular parts of the climb, but can then show that they are able to finish the remainder of the climb without assistance. This is the conceptual framework of the integrated testlet. Consider CR8 (Figs. \ref{fig:CR_and_IT_examples}(b) and \ref{fig:testlet_maps}(b)) which deals with rotational dynamics, frictional torque, and work. In the first part of the question, students are asked to solve for the speed of a falling object that is tied to a frictionally-coupled rotating cylinder. In the second part, the students are asked about the work done by the friction in the cylinder as the object falls. As shown in Fig. \ref{fig:testlet_maps}(b), there are two conceptually-distinct ways to solve CR8(b); the less-efficient method involving the solution to CR8(a). When constructing IT8, IT8-i is a required and important intermediate to CR8(a), with IT8-ii being identical to CR8(a). Then IT8-iii tests a seemingly non-integrated step that is in fact meant to represent exactly the kind of scaffolding motivated by Ding et al. \cite{Ding_PRSTPER_2009}. Finally, IT8-iv is equivalent to CR8(b), thus allowing IT8 and CR8 to test the same conceptual domain. As shown in Table \ref{tab:summary}, CR8 proves to be the second most difficult of all of the exam questions in the course, and IT8 is the most difficult IT given in the course. Thus, the cuing and scaffold-building provided by intermediate steps IT8-i and IT8-iii do not significantly simplify the problem, as the IT difficulty value ($p'$) is still below that suggested by Fig. \ref{fig:difficulty_correlation}. Without direct instructional cuing of how the solution to IT8-iii can help solve IT8-iv, students must still demonstrate that they know how the questions are linked; they must demonstrate the integrated conceptual understanding that is being tested. This notion is further confirmed by the very high value of $r'=0.63$ for IT8.  
	All eight integrated testlets in our study were created with similar considerations to those outlined above. We considered which steps in the solution to the matching CR question we anticipate will be most difficult and then decided whether to add an intermediate step as an item within the IT. As with IT8, if the solution for a question draws on concepts from different parts of the course we use a mid-testlet question to provide subtle cuing and scaffolding. Because of the aims of the current study we always made sure that the final testlet item was identical to the final CR sub-question. However, because solving an IT may take less time than solving an equivalent CR, and furthermore because test-takers have confirmatory or corrective feedback at every step, it is certainly possible for an IT to ask questions beyond the scope of the CR, and to do so in a similar time-frame on an exam. Thus ITs could ultimately assess deeper knowledge (i.e. climb a higher mountain) than is viable with a CR question.

\bibliography{Slepkov_IF-AT}{}

\begin{thebibliography}{}
\bibitem{MultipleKeys}
There is, however, a variety of IF-AT forms with different answer keys
  available from Epstein Educational---a fact that greatly aids in test
  security.
\bibitem{WilcoxonDifficulty}
The difference in difficulty between matching IT and CR questions is
  statistically significant. Further, a Wilcoxon signed-rank test shows that
  there is a large effect size to this difference ($W=36.0$; $Z=2.5$;
  $p=0.012$, $ r=0.63$). With only 8 pairs of data, such a non-parametric test
  is most appropriate. Nonetheless, a paired-sample t-test also yields
  $p<0.05$).
\bibitem{WilcoxonDiscrimination}
With only 8 pairs of data, there is insufficient statistical power to establish
  statistical significance in the discrimination parameters of CR questions and
  their matching ITs. A Wilcoxon signed-rank test yields $W=7.0$; $Z=1.54$;
  $p=0.12$, $ r=-0.385$, and we cannot reject the null hypothesis. Note that
  this finding is also supported by a paired-sample t-test, which yields
  $p>0.05$.
\bibitem{testlet_boost}
In every one of the more than 50 testlets we have analyzed over the past few
  years we have found that the testlet-level $r'$ is higher than the average of
  the $r'$ values of the comprising MC items. We have not been able to find
  this discussed in the classic testlet theory literature and are currently
  studying this further.
\bibitem{average-discrimination}
Since this measure is being reported for direct comparison with that found in
  other MC exams, we calculated this value by only including multiple choice
  items in our midterms and finals. That is, we removed all CR items and
  calculated the item-total pearson-$r$ for the 8 MC items of each midterm and
  the 12 MC items of each final, then we averaged all 40 values.
\end{thebibliography}
\bibliographystyle{prst-per}


\end{document}